# The 1908 Tunguska event: analysis of eyewitness accounts of luminous phenomena collected in 1908

Andrei Ol'khovatov

https://orcid.org/0000-0002-6043-9205
(Retired physicist)
Russia, Moscow
email: olkhov@mail.ru

**Dedicated to the blessed memory of my grandmother ( Tuzlukova Anna Ivanovna ) and my mother ( Ol'khovatova Olga Leonidovna )**

**Abstract:** Historically there were two main reasons to assign the 1908 Tunguska event to a spacebody infall: a) newspaper notes about a fall of a meteorite near the town of Kansk (later claimed to be false); b) eyewitnesses reports about seeing luminous phenomena in the sky. This paper examines accounts of the Siberian eyewitnesses about luminous phenomena in the sky, collected in 1908, immediately after the event. The conducted generalization of the available accounts reported in 1908 indicates that eyewitnesses reported several types of luminous phenomena.

## 1. Introduction

More than a century has past after the famous 1908 Tunguska event in Siberia (abbr. Tunguska). But till now there is no solid scientific proof of any of the numerous hypotheses proposed.

In the morning of June 30, 1908, thunderous sounds were heard by population north and northwest of Lake Baikal in Central Siberia. In some places also the ground trembled. Reports of luminous phenomena in the sky came from various points of the region. Soon a newspaper story appeared about a fall of a large meteorite near the town of Kansk, but then it was recognized as wrong.

This paper examines accounts of Siberian eyewitnesses about luminous phenomena in the sky, collected in 1908, immediately 'in hot pursuit'.

Translation from Russian is done by the author of this paper (i.e. A.O.), unless otherwise is stated. The translation was done as close to originals as possible. In some places of translations an alternative translation and/or comments in {...} are given by the author of this paper for better understanding. Please pay attention that the Tunguska event occurred on June 30, 1908 (on the Gregorian calendar) or on June 17, 1908 on the Julian calendar which was widely used in Russia in 1908.

On Fig.1 there is a map of the region (based on English edition of [Fesenkov, 1966] with some additions). Red arrows drawn by the author (A.O.) 1, 2, 3 are proposed (by various researchers) trajectories of the alleged Tunguska spacebody (see for details later in the paper).

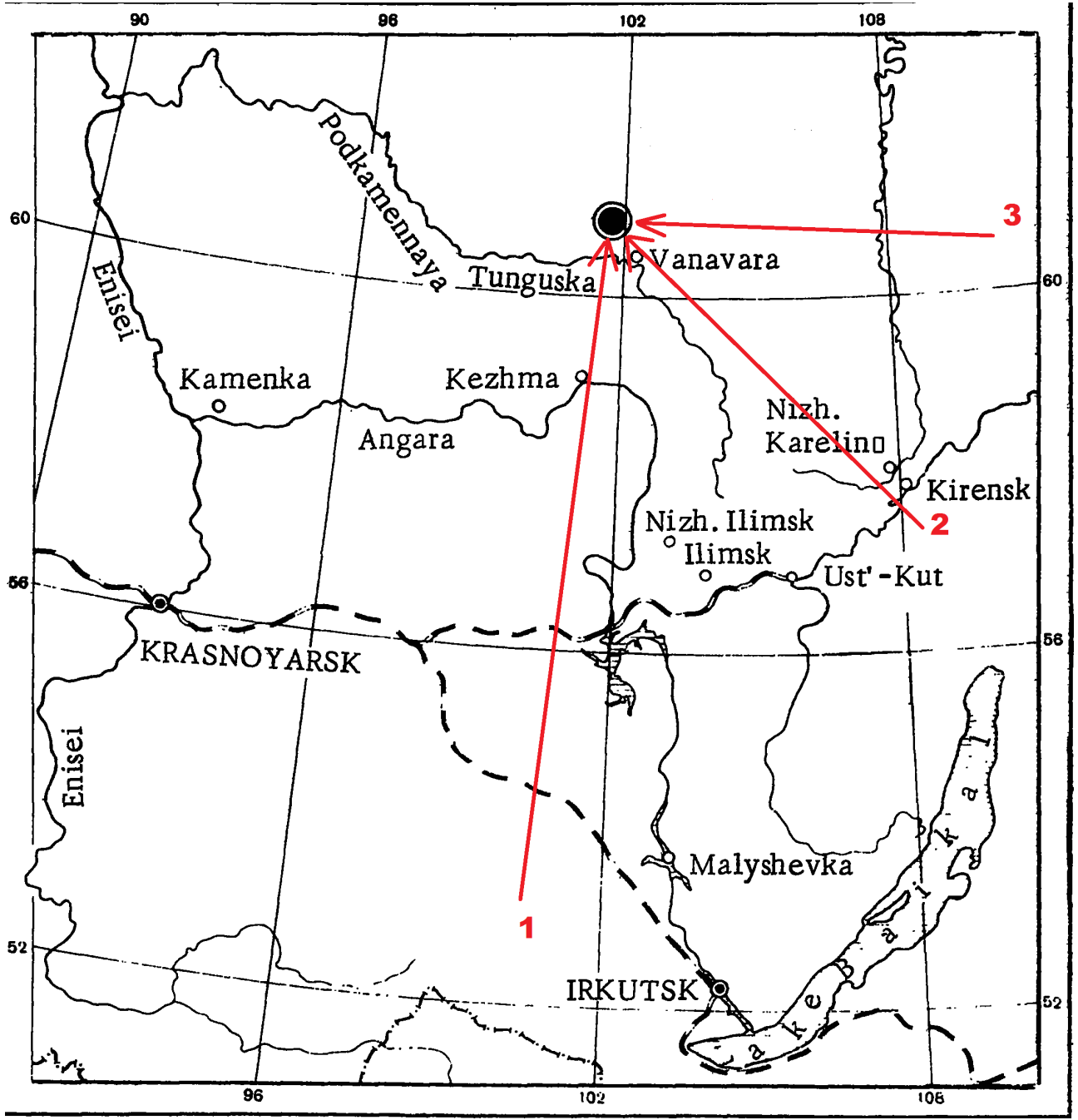

**Fig.1**

For better understanding positions of settlements are given. Please pay attention that so called the epicenter of the Tunguska event is assigned to 60°53' N, 101°54' E. The positions are taken from [Vasil'ev et al., 1981], and [Razin, 2008], and maps. Please pay attention that the settlement's names (especially of small settlements) in various publications can differ a little bit (as, for example, N-Karelino, Nizhnekarelina, etc.). Anyway here they are:

Ancir 56°15' N; 95°31' E
Bodaibo 57°51' N; 114°11' E
Bel'skoe (from a map) 57° 49' N; 92° 10' E
Bratskoe 56°21' N; 101°55' E
Bur 58°49' N; 107°00' E
Chechuyskoe 58°04' N; 108°42' E
Chelpanovo 54°06' N; 105°34' E
Chirida 58°57' N; 101°28' E
Dalai (Dalay) (taken from a map) 56°18' N; 95°54' N
(the) epicenter (of the Tunguska event) 60°53' N; 101°54' E
Filimonovo 56°12' N; 95°28' E
Ilimsk 56°46' N; 103°52' E
Irkutsk 52°17' N; 104°18' E
Kamenskoe 58°21' N; 92°28' E
Kansk 56°12' N; 95°42' E
Karapchanskoe 57°54' N; 102°49' E
Kezhemskoe (Kezhma) 58°58' N; 101°07' E
Kirensk 57°47' N; 108°07' E
Korelino (Karelino, Karelina, see below) 57°49' N?; 107°26' E?
Malyshevka 53°44' N; 103°24' E
Mukhtui 60°44' N; 114°56' E
(the) Mursk rapids 58°27' N; 98°30' E
Mutinskaya 58°38' N; 110°01' E
Nizhne-Ilimskoe (N.-Ilimsk) 57°11' N; 103°16' E
Nizhne-Karelino (N-Karelinskoe, Nizhne-Karelina, etc.) (from a map) 57°58'
N; 107°49' E
Podkamenskoe 57°58' N; 108°31' E
Preobrazhenka 60°01' N; 108°05' E
Shamanskie water-measuring posts 55°36' N; 101°55' E
Teterya (Tatere) 60°09' N; 102°14' E
Tuba 57°39' N; 103°23' E
Ust-Kut 56°46' N; 105°39' E
Vanavara 60°20' N; 102°17' E
Vitim 59°27' N; 112°33' E
Voronino (Voronina) 57°47' N; 108°05' E (from a map)
Yarskaya 57°09' N; 103°22' E
Yeniseysk 58°27' N; 92°10' E
Zavalomnoe 58°00' N; 108°40' E
Znamenskoe 54°42' N; 104°50' E

Sometimes there is confusion due to similar geographic names. For example,

the author failed to find Korelino (Karelino, Karelina) on a map, but discovered instead Verkhne-Karelina (57°52' N, 107°46' E) which is indeed about 20 versts (1 verst = 1.0668 km) from Kirensk (see an account by G.K. Kulesh below).

## 2. The accounts

First reports in mass-media about luminous phenomena appeared in 1908 in 'hot pursuit'.

Probably the first newspaper story which described luminous phenomena was an article in the "Sibir" newspaper (published in Irkutsk). The article is by the newspaper correspondent (issue of July 2 on the Julian calendar). A fragment of the article is shown on Fig.2.

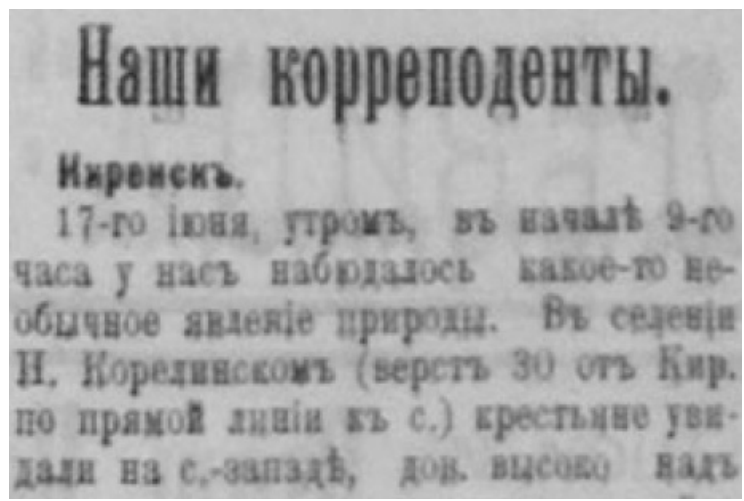

**Fig.2**

Please pay attention that expressions like "versts 30", "versts in 6", "versts in 180", "hours p.m. 2" often have a tint in Russian as approximate, i.e. "about 30 versts", "about 6 versts", etc.. They are left in translation to be as close to original as possible. Here is the translation:

"**Kirensk.**
On June 17th in the morning, at the beginning of the 9th morning, it was observed some unusual phenomenon of nature here. In the village of N-Karelinskoe (versts 30 from Kirensk to the north), the peasants saw in the northwest, quite high above the horizon, some extremely strong (it was impossible to look at) a body shining with a white bluish light, moving for 10 minutes from top to bottom. The body was represented in the form of a "pipe", i.e. cylindrical. The sky was cloudless, only low above the horizon, in the same direction in which the luminous body was observed, a small dark cloud was noticeable. It was hot and dry. Approaching the ground (a forest), the shiny body seemed to spread out, a huge puff of black smoke formed in its place and an extremely strong knock (not thunder) was heard, as if from large falling stones or cannon fire. All the buildings were shaking. At the same time, a flame of indeterminate shape began to burst out of the cloud.

All the inhabitants of the village in panic fear ran down a street, the women were crying, everyone thought that the end of the world was coming... In the end, they decided to send a messenger to Kirensk to find out what the phenomenon that frightened them so much means (the above information was obtained from this messenger).

The writer of these lines was at that time in the forest, versts in 6 from Kirensk to the north and heard in the northwest, as it were, cannon fire, repeated intermittently for 15 minutes several (at least 10) times. In Kirensk, in some houses, in the walls, facing northwest, the windows rattled. These sounds, as it turned out now, were heard in the villages of Podkamenskoe, Chechuyskoe, Zavalomnoe and even at the station Mutinskaya, versts in 180 from Kirensk to the north.

At that time, in Kirensk, some observed as a fiery red ball in the northwest, moving, according to some, horizontally, and according to others, very obliquely. Near Chechuyskoe, a peasant driving through a field observed the same thing in the northwest.

Near Kirensk in the village of Voronino, peasants saw a fiery ball, fallen to the south-east of them (i.e. in the direction opposite to the one where N-Karelinskoe is located).

The phenomenon has aroused a lot of talks. Some say that it is a huge meteorite, others - that it is a ball lightning (or a whole series of them).

Hours p.m. 2  between Kirensk and N-Karelinskoe (closer to Kirensk) on the same day there was an ordinary thunderstorm with heavy rain and hail."

A newspaper  "Krasnoyarets" in the issue of July 13, 1908 (the Julian calendar) published the following article  (scan of its beginning is on Fig.3, translation is

below):

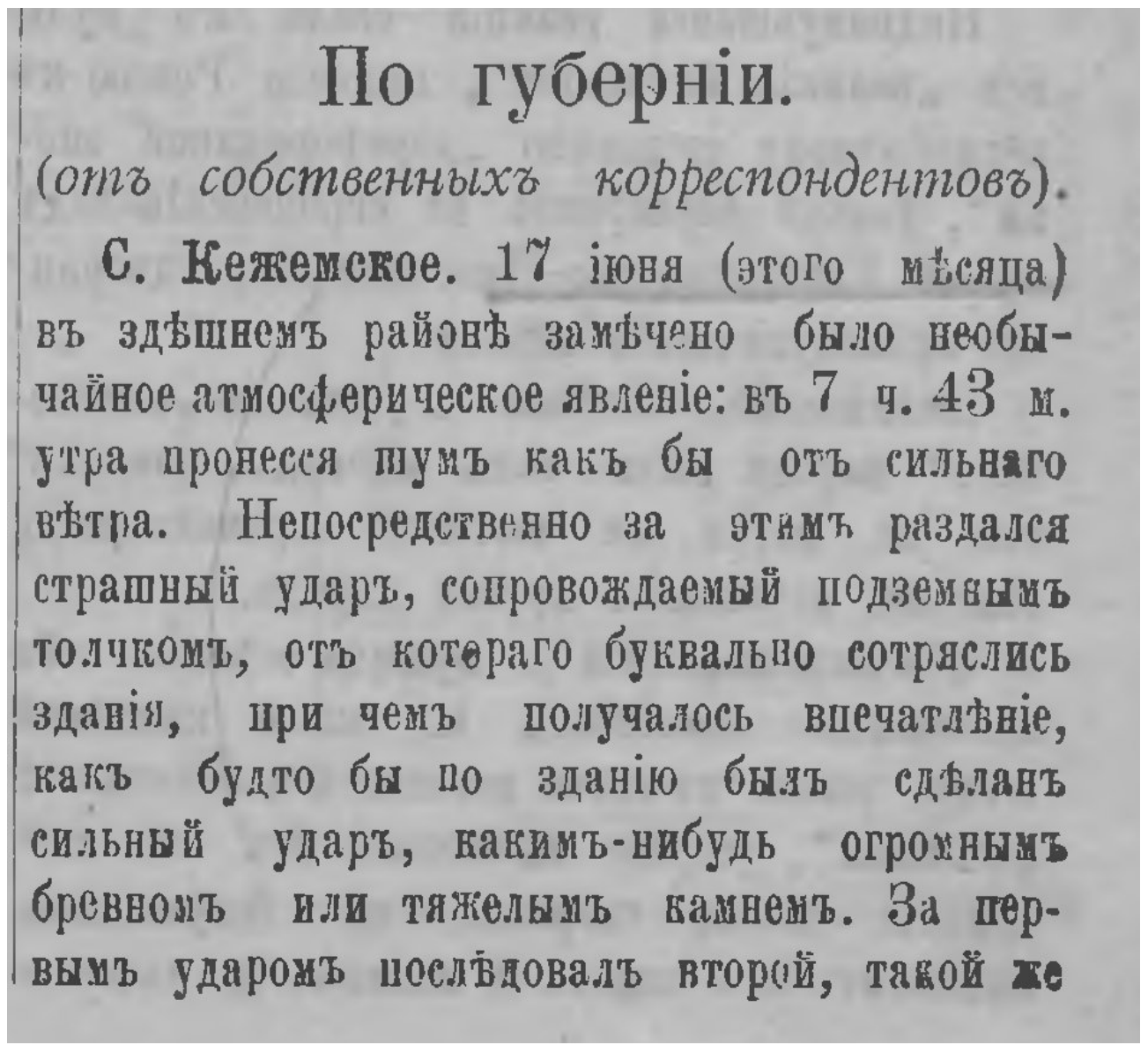

**Fig.3**

"S. {settlement - A.O.} Kezhemskoe. On June 17th (this month), an extraordinary atmospheric phenomenon was noticed in this region. At 7:43 a.m., a noise as if from a strong wind swept by. Immediately after that, a terrible blow was heard, accompanied by an underground push, from which the buildings literally shook, with which the impression turned out, as if by the building was hit hard by some huge log or heavy stone. The first blow was followed by a second, of the same strength, and a third. Then - the time gap between the first and third blow was accompanied by an extraordinary underground rumble, similar to the sound from the rails, along which ten {in the source - "desyatok", which often means "about ten" -A.O.} trains seemed to pass at the same time. And then for 5-6 minutes there was exactly as artillery firing: about 50-60 blows followed after {meaning "separated" - AO.} short and almost identical intervals of time, gradually, the blows became weaker towards the end. Following a 1.5 - 2 minute break (after the end of the continuous "shooting"), six more blows were heard one after another, like distant cannon shots, but still clearly audible and felt {"cleaned" - in the original, probably a typo - A.O.} by shaking the earth.

The sky, at first glance, was completely clear. There was no wind or clouds. But with close observation, in the north, i.e., where the blows seemed to be heard - on the horizon, something like an ash-colored cloud was clearly noticed, which gradually decreasing, it became more transparent, and by 2-3 o'clock in the afternoon it completely disappeared.

According to the information received so far, the same phenomenon has been observed in the surrounding villages of Angara at a distance of 300 versts (down and up) with the same force. There have been cases that from the shaking of the houses shattered the glass in the casement frames. How strong were the first blows, it can be judged by the fact that in some cases horses and people fell from their feet.

According to eyewitnesses, before the first blows began, some kind of celestial body of a fiery kind cut through the sky from the south to the north with a tendency to NE, but for the speed (and most importantly - the surprise) of the flight, neither the size nor the shape of it could be seen. But on the other hand, many and in various villages perfectly saw that with the touch of the flying object to the horizon, in the place where abovementioned peculiar cloud was subsequently noticed later, but much lower than the location of the latter - at the level of forest tops, it was as if a huge flame broke out, splitting the sky. The radiance was so strong that it was reflected in rooms with windows facing north, which was observed, by the way, by watchmen of the volost board. The glow lasted, apparently, at least a minute, so as it was noticed by many peasants who were on arable land. As soon as the "flame" disappeared, at once the blows started.

With the ominous silence in the air, it was felt that some extraordinary phenomenon is happening in nature. On the island opposite the settlement, horses and cows began screaming and running from edge to edge. It got the impression that the earth is about to open up and everything will fall into the abyss. Terrible blows were heard from somewhere, shaking the air, and the invisibility of the source inspired some kind of superstitious fear. Literally it was "dumbfounded." "

The last part of the article is omitted as it does not contain new description of the phenomenon.

There was an article "A meteor, a lightning or an earthquake" in the July 15, 1908 (in the Julian calendar) issue of the newspaper "Golos Tomska" which included the following text (translated):

"For some time, rumors were spreading that the aerolite fell near the village of Dalai and as if many had seen how it flew, and that during the fall, this aerolite hit a tree - a thick pine, which it smashed and there were a lot of such stories."

The village Dalai (its official name was: Dalaiskaya) was about 19 km to NE of Kansk. The article also repeats stories from early publications on the matter. A fragment of the article is shown on Fig.4.

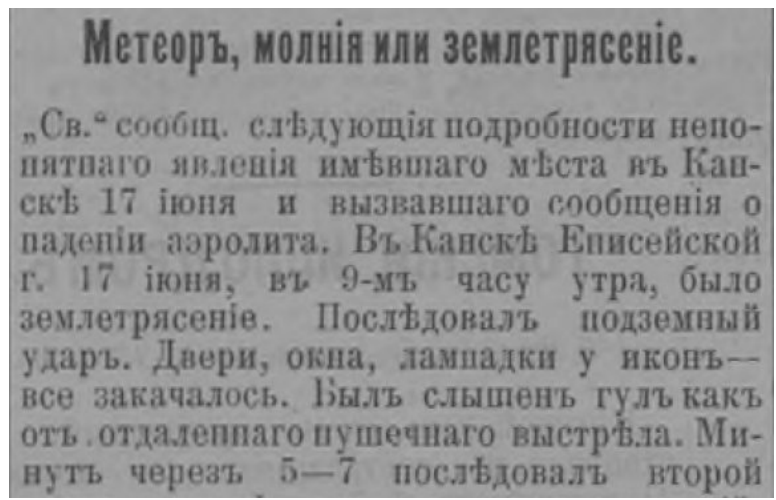

**Fig.4**

One of the major sources of the 1908-witness's accounts were accounts collected by A.V. Voznesenskii who in 1908 was a director of the Magnetic-meteorological observatory in Irkutsk.  A.V. Voznesenskii had a large network of correspondents in the region who regularly reported him about earthquakes. Voznesenskii got  ~60 reports about the 'Tunguska earthquake' in total. It is very important that the reports were delivered soon after the event and by rather experienced observers.

Voznesenskii sent out 3 types of questionnaires about earthquakes (questionnaires: No.2, No.3, No.4). Let's call them as: questionnaire -A, -B, -C respectively. Only one reply to the questionnaire-C is known, so it is omitted here as it does not contain any info on luminous phenomena.

The questionnaire-A contains questions regarding: a place of observation (1); a date and time of the observation (2); details of the earthquake and whether it was accompanied with subterranean rumble (3); about possible damage and other effects of the earthquake (4); about an intensity of earthquake (5); surname and address of the observer (6). Dates in the accounts/replies are in the Julian calendar unless otherwise is stated. Here are translations of collected replies on the questionnaire-A which mentioned luminous phenomena (taken from [Vasil'ev et al., 1981], translated):

"L.D. Klykov, head of the post office in the village of Znamenskoe, Irkutsk province, July 11, 1908.
1. In the settlement of Znamenskoe, Irkutsk province.
2. On June 17, 1908, on Tuesday, approximately around 8 o'clock in the morning, the clock was checked with the Zhiganovskii telegraph.
3. The underground rumble was not heard... There were two shocks, 3-4 seconds apart from each other, the whole earthquake lasted 7-8 seconds.
4. There was no. At the same time, the fall of the aerolite was noticed: approximately to the southwest of S. {settlement –A.O.} Znamenskoe; a fiery strip was visible, the sky was completely clear, a strong thunder was heard at the end with an explosion."

On the web-site tunguska.tsc.ru in the point 3 (after the word "heard"), there is also a phrase (by Klykov) stating intensity of the earthquake as I=5 (Rossi-Forel scale), and also point 1 is added from the web-site.

The next reply is from Vakulin - a head of the Nizhne-Ilimskoe post-office which was written as a letter of July 28 ([Vasil'ev et al., 1981], translated):

"On Tuesday, June 17, at about 8 o'clock in the morning (the clock has not been checked), according to the stories of a large circle of local residents, they were initially seen in the north-west direction a fireball descending indirectly {obliquely, gently sloping? – A.O} along the horizon from east to west, which, when approaching the earth, turned into a pillar of fire and instantly disappeared; after disappearing in this direction, a puff of smoke was visible rising up from the earth.

A few minutes later there was a strong noise in the air with muffled individual blows like thunderclaps. These blows were followed by 8 strong blows, similar to gun shots. The last blow was with a whistle and especially strong, from which the surface of the earth and buildings slightly oscillated. According to the collected information, these phenomena are confirmed by the residents of the terminal {closest to a border – A.O.} villages of the Nizhne-Ilimskaya volost, part of the Kocheng and Karapchansky volosts, located from east to west, on a space of about 300 versts"

From the same place (Nizhne-Ilimskoe) there was another report. Kokoulin (an agronomist from this village) in a letter dated July 25, 1908 informed Voznesenskii ([Vasil'ev et al., 1981], translated):

"On June 17, at about 7:15 a.m., workers building the bell tower saw a fire block {log, chunk, lump – A.O.} flying, it seems, from the southeast to the northwest. First there were two thunders (like gunfire), then a very strong thunder with a concussion. More thunders were heard. The earth shaking was noticed. One girl fell from a bench (a priest's servant). The population was frightened. This fireball (an obscure word in the original, like "meteor") was seen in Karapchanskoe and were heard the blows. It was a clear day, and therefore thunder bewildered the audience. In Nizhne-ilimskoe, two Tunguses told that the meteor formed a lake by falling, which boiled for two days. The Tunguses were ready to show this lake, but no one believed this story."

There is one more letter to Voznesenskii from Kokoulin of Sept.14, 1908 ([Vasil'ev et al., 1981], translated):

"It was possible to more accurately delineate the area of propagation of the sounds accompanying the phenomenon. It turns out that the rumble was heard near Verkholensk (in the village of Chelpanovo) - on the one hand and in Mukhtui - on the other, i.e. at a distance of about 1300 versts along the Lena (river - A.O.). Aside from Lena, the phenomenon was more or less thoroughly observed in Nizhne-Ilimsk. Eyewitnesses say that at the place of the fall of the body (or, perhaps more precisely, at the place of its moving below the horizon) clouds of black smoke were rising in a column. The Tunguses, who wandered behind the Nizhne-Karelinskoe village (to WNW from Kirensk), say that the thunder was terrible, but deer did not show the slightest excitement, as during a normal thunderstorm. The earthquake in its usual form was not observed, it was good (in some places even too good) noticeable only the trembling of buildings from the roar. The

meteor was moving from east-southeast to west-northwest."

Voznesenskii also spread the more detailed questionnaire (questionnaire-B) about the June 17 earthquake including the questions ([Vasil'ev et al., 1981], translated):
:
1. A place of observation;
2. A time of the beginning and the end of the earthquake;
3. Directions;
4. Details of the quaking;
5. Was a rumble heard?
6. Personal impressions.
7. Was furniture, utensils moved?
8. Here Voznesenskii asked to evaluate the earthquake by Rossi-Forel scale and explained the scale.

M.R. Romanov, a peasant from Nizhne-Ilimskoe, gave the following answers to the questions of questionnaire-B ([Vasil'ev et al., 1981], translated):

"1. Nizhne-Ilimskoe.
2. No one observed the time of the beginning and the end of the earthquake, and therefore it is impossible to determine exactly, it was approximately at the beginning of the 9th hour in the morning.
3. –
4. Oscillation with concussion.
5. Two rumbles were heard, similar to a cannon shot.
6. Fright.
7. It was not noticed, but there was a tremor of glass in the windows.
8. Against a score of three, the word "majority" is underlined.
    At the end of the questionnaire M.R. Romanov wrote: "At the beginning of the 9th hour of the morning local time, a fiery ball appeared, which flew in the direction from the southeast to the northwest. This ball, approaching the earth, took the form of a flattened ball from above and below (as it was visible to the eye); approaching even closer to the earth, this ball had the appearance of two pillars of fire. When this huge mass fell to the ground, two strong, thunder-like impacts occurred, and as a proof that it was not a thunder, it can be that the sky was completely cloudless, then a noise was heard, as if from a strong wind; the duration of this phenomenon is about 15 minutes.""

The following text was published in the book [Krinov, 1949]:

"Another very interesting description of the meteorite fall, printed in newspapers, was received from the village of Nizhne-Ilimskoe. The author of the article, N. Ponomarev, reported that the population of Nizhne-Ilimskoe and surrounding villages was alarmed by an unusual phenomenon on June 17. At 7.20 a.m. , a strong noise was heard over Nizhne-Ilimskoye, which turned into thunderclaps. Meanwhile, the sky was cloudless. Some houses were shaken by the thunders. Many residents saw that before the thunder, "some kind of fiery body like a log" rapidly swept over the earth from the south to the northwest. Immediately after that, a thunder was heard, and in the place where the fiery body disappeared, "fire" appeared, and then "smoke"."

The author of this paper failed to find this story in newspapers. It looks like authors of [Vasil'ev, et al., 1981] also failed as they just cited the book, and then added a quote from [Voznesenskii, 1925] which is much more detailed. Here it is ([Vasil'ev, et al., 1981], translated):

"In the section of newspaper materials, N. Ponomarev's message from Nizhne-Ilimsk already was presented. A.V. Voznesenskii in his article [6] quotes the testimony of N. Ponomarev, accompanying them with the following words: "Finally, we have the same data in the correspondence of N. Ponomarev from Nizhne-Ilimsk." It is unclear whether this is a reprint from another some newspaper, or a letter from Ponomarev to Voznesensky. Since it is much more detailed than the above, we will quote it here on [6]:

"The population of Nizhne-Ilimsk and surrounding villages is alarmed today (June 17) by an extraordinary phenomenon. At 7:20 a.m. over Nizhne-Ilimsk, during a very good weather condition (the sky was covered with clouds in some places), an out-of-the-ordinary noise approaching the village was heard, as it approached, turned into thunderclaps. After the rumblings, the strongest impact was heard throughout the district, which almost caused panic in the population. I slept. When thunderclaps were heard, I woke up, and at the moment of the impact I felt my house shake, dishes rattled in the kitchen, and a chair that was standing tightly against the wall was pushed back for a vershok {i.e. about 4.5 cm -A.O.} to the middle {of the room - A.O.} by the oscillation of the wall, the servants, who happened to be standing on the bench, almost fell from the concussion. Hastily dressed, I run out into the street, I read a big alarm on their faces, in some places the population has climbed onto the roofs of houses and is looking in the direction where the impact was heard. One {person – A.O.} comes up to me and informs me that he saw (before the thunderclaps) as a fiery body like a log rapidly rushed over the ground and disappeared, then immediately a thunder was heard. The same was reported by another person who saw it together with the contractor. A guy rides a horse and reports that he also saw some kind of fiery body, saw how in the place where it fell, fire appeared first, and then (when the impact occurred) instead of fire smoke appeared. My father and two brothers were about 6 versts from Nizhne-Ilimsk on a fishing trip and clearly heard - how before the strong impact {it looks like Ponomarev is talking about the strongest impact mentioned early – A.O.} there were two thunderclaps, not so strong, and after the impact - less strong thunders were heard a lot - up to 100 and in different places in three directions. One of the brothers, who was in a war, compares what happened with that the moment when enemy's firing opens and the big military guns rattle... The fiery body was seen also by another person who drove up to us from the village 7 versts away. The body swept from the south to the northwest, and all those who saw it unanimously confirm this, accurately indicating the direction.""

Here is an important report to Voznesenskii from an observer of a meteorological station ([Vasil'ev et al., 1981], translated):

"The observer of the meteorological station in Ilimsk, N.N. Polyuzhinskii, in a letter dated June 21, 1908, reported:
"June 17, 1908 at 8.30 a.m. in the morning, there was a strong noise and a sound like a strong thunderstorm and cannon shots, following one after another (drummroll) {like drumming? - A.O.}, probably from a passing meteor (aerolite)."

A month later, the same observer sent A.V. Voznesensky a new letter in which he reported additional details ([Vasil'ev et al., 1981], translated):
:

""June 30, new calendar art. (June 17, old calendar), Tuesday, about 8 o'clock in the morning, with clouds of $10^0$, a thunder was heard in the air in the south-southeast side, like rapid revolver shots, then the thunder became stronger and stronger, like strong powder explosions and cannon shots, which, as they approached Ilimsk, turned into a terrible crackling, so that a small earthquake turned out (the swinging of the lamp in front of icons and a push noticed by people sitting quietly). After the crackling (in the air), a noise (rumble) was heard, and the thunder began to move away in a north-north-westerly direction; the thunder lasted about 20 minutes, and there was no lightning.
During the thunder, one philistine of Ilimsk was 4 versts from Ilimsk up the Ilim River and saw a "flying star with a fiery tail" that fell into the water, and its tail disappeared into the air.

In the village of Yarskaya (about 60 versts from Ilimsk down the Ilim River), three women saw a "fiery ball" (flying); it is unknown where it disappeared, as the women were afraid of it and ran away from the field home.

In my first message about the same, the time of the thunder phenomenon is indicated incorrectly, by a mistake. N. Polyuzhinskii".

In addition, N.N. Polyuzhinskii gave the following answers to the questions of questionnaire-B:
June 17 , 1908 Tuesday.
1. The town of Ilimsk, Irkutsk province, the Kirensky district.
2. At 8 o'clock in the morning, two minutes (approximately).
3. SE - NNW (approximately).
4. In the form of a single push.
5. it was.
6. Under the ground as if stones were falling.
7. There was no.
8. Against "score 2" the words "some persons at rest" are underlined"

Head of the Kirensk Meteorological Station G.K. Kulesh in a letter dated June 23 collected and reported accounts ([Vasil'ev, et al., 1981], translated, sazhen = 2m 13cm):

"On June 17 (old calendar), a phenomenon was observed on the NW from Kirensk, which lasted from about 7 h. 15 min. to 8 h. in the morning. I did not get to observe it, since I, after recording the meteorological instruments, sat down to work. I heard muffled sounds, but I took them for volleys of gunfire on the military field beyond the Kirenga River. When I finished my work, I looked at the barograph tape and to my surprise noticed a line next to the line made at 7 o'clock in the morning. This surprised me, because during the continuation of work I did not get up from my seat, the whole family was asleep, and no one entered the room. Here's what happened (I'm passing on the essence of the eyewitness stories). At 7.15 am a pillar of fire appeared on NW, four sazhens in diameter, in the form of a spear. When the pillar disappeared, five strong short thunders were heard, as from a cannon, quickly and clearly following one another; then a thick cloud appeared in this place. After 15 minutes, the same thunders were heard again, after another 15 minutes, it was also repeated. A ferryman, a former soldier and a man in general experienced and developed, counted 14 thunders. According to the duty of his service, he was on the shore and observed the whole phenomenon from the beginning to the end. The pillar of fire was visible to many, but the thunders were heard by even more people. The peasants of the villages closest to the town come to the town and ask: "What was that? Does this portend war?" There were (in the town) also peasants from the village of Karelino {possibly Verkhne-Karelina or Nizhne-Karelino – A.O.}, which lies 20 versts from Kirensk on the nearest Tunguska {apparently, the Nizhnyaya Tunguska river –A.O.}, they reported that they had a strong shaking of the soil, so that windows were broken in the houses.

From other sources, it is reported that a lake was formed in the mountains seven versts from the village of Karelino {possibly Verkhne-Karelina or Nizhne-Karelino – A.O.}. According to the stories of the peasants there was a flat place, swampy. In summer and winter, some vapors rose from the swamp. This swamp has become a lake. These stories are not verified. Now this phenomenon has given rise to a lot of the most fantastic stories and assumptions among the people... It has been correctly established that a meteor of very huge dimensions fell, because in perfectly clear sunny weather the column seemed to be in four sazhens in diameter. A grey-colored cloud was seen, and then turned in the dark; there were thunders, the number 14 in three steps; there was oscillation of the soil; the line on the barograph tape serves as proof of this. In addition, a contractor Yashin lives in the neighborhood of the progymnasium. He was in the yard when a board leaning against the fence fell, although it was completely quiet in the yard. Or maybe there was a strong shaking of the air, because the last thunders were the strongest.

According to the story of one inhabitant of Kirensk, at that time he went to a chest for something; he had just opened it when thunders started, and he was swayed towards the chest, as if from a strong wind."

He later adds {probably in another letter? – A.O.}: "I received new information about the meteor from the students. Two pupils told me that they were Burskie {residents of the Bur -  A.O.} (the village of Bur on the Nepa River, a tributary of the Tunguska)  the peasants heard from the Tungus - eyewitnesses of the meteor fall such a picture: "When the meteor fell, thick smoke rose, thunders were heard, peat and forest caught fire, so the Tunguses extinguished the fire for three days.""

In August there was another letter by Kulesh ([Vasil'ev, et al., 1981], translated):

"G.K. Kulesh also filled out the questionnaire No.3 {i.e. -B – A.O.}, where he signed himself as a "town school teacher". Apparently, he conducted meteorological observations along.
1. In houses and in the open place by many persons.
2. About 8 hours in the morning and lasted until 9 o'clock.
3. NW
4. The trembling of the glasses in the windows was heard.
5. Thunders were heard, like cannon shots, repeated three times.
6. –
7. –
8. Underlined 2 points.

A postscript is given to the questionnaire:
"A lot has already been reported to the Observatory about the former earthquake, although many eyewitnesses along Lena {river – A.O.} did not pay attention to the earthquake and did not notice it, being surprised unusually strong thunders. Now it turned out that the thunders were heard in areas very far from each other: there is information, quite reliable, that the thunders were on Bodaibo, in Vitim and up {upstream- A.O.} Lena to Ust-Kut, in Nizhne-Ilimsk; in Nizhne-Ilimsk, the direction of thunders is village Tuba on the Ilim river. The pillar of fire is visible to many, its shape in the form of a spear is also identified. Smoke or a gray cloud, which then turns into a dark one, is also noticed by many. I could not establish when there was a trembling of the windows in the houses, during the thunders, before or after them. The strongest blows were the last, there was a strong concussion of the air. Stories about what is on the near Tunguska {apparently, Nizhnyaya Tunguska river –A.O.} and the village Korelino {possibly Verkhne-Karelina or Nizhne-Karelino – A.O.} formed a lake, turned out to be wrong. The peasants of this village were so stunned by the thunders that they sent a deputation to the town to the local archpriest to ask if the end of the world was beginning, how they are preparing for it in Kirensk. That there was a shaking of the ground, I could conclude from the fact that the barograph marked the line on the tape, and I firmly remember that no stranger entered the room, and I did not get up from my seat, I could not push the device. The thunders were heard by me, but since the windows were closed from NW, and only open on S, then I took the thunders for volleys of rifle shots on a military field.""

A couple of words about G.K. Kulesh. The article by V. Vlasov ("Meteorologicheskii Vestnik", N1 for 1912) gives an excellent review of the work of the Kirensk meteorological station, which for over 20 years (the Vlasov's visit to it in 1911 is described) has provided reliable and valuable material, "thanks to the rare attention and diligence of the town school inspector G.K. Kulesh, who has been receiving until now for his work one hundred rubles a year." Remarkably that Kulesh

who investigated the event did not mention about super-bright bolide…

The next report is from rather remote area ([Vasil'ev, et al., 1981], translated):

"A correspondent from the village of Malyshevka, located near the railway, at a distance of about 700 km from the place of the meteorite fall, hereditary honorary citizen I.V. Nikolskii gave the following answers to questionnaire No. 3 {i.e. -B - A.O.}:
2. 8 hours 15 minutes of the morning.
3. To the northeast.
4. –
5. A dull thunder, as if from a carriage that had moved down on the pontoon's scaffold.
6-7. –
8. score 2.
At the bottom of the questionnaire there is a postscript:
"Exactly at 8 o'clock 15 min in the morning, a boy working in the volost's {i.e. parish - A.O.} yard saw a fire falling in the form of a stump (according to him, in the form of a bucket) in the direction of the northeast; the same thing and in the same direction was seen by some workers who were in the forest about 20 versts from Malyshevka. I personally did not see the fire, but I and some others heard the dull thunder.""

The observer (T. Grechin) of the Shamanskie water-measuring posts wrote to Voznesenskii on June 18, i.e. on the next day after the event ([Vasil'ev, et al., 1981], translated):

"On June 17, at 8 o'clock in the morning, there were some strong thunders several times in the north side like thunderclaps {i.e. as during a thunderstorm – A.O.}, from which the windows trembled in the frames, the trees bent down and the leaves of them shook; at the same time it was clear and quiet, and even the water did not lose its gloss, no damage was noticeable. As local peasants who were at field work at that time told me, they saw some kind of a fiery ball flying in the north side, from which such strong thunders like explosions seemed to occur."

The next account is from rather remote area ([Vasil'ev, et al., 1981], translated):

"A. Goloshchekin from the village of Kamenskoe, located on the Yenisei River (near Yeniseisk), 600 km west-southwest of the meteorite impact site, reported in his letter dated June 30, 1908 that "at 7 o'clock in the morning the following phenomenon was observed in the village of Kamenskoe: three underground thunderclaps were heard in the direction from northwest one after the other. At the same time, some observed a concussion. From the inquiries of local residents, I knew that a few minutes earlier, some of them had seen a body (as if detached from the sun) more than a yard long, oblong in shape and tapering towards one end; his head was as light as the sun, and the rest was of more hazy color. This body, having flown through space, fell in the northeast.""

. In 1925 Voznesenskii published an article [Voznesenskii, 1925] in which, in particular, he wrote (translated):

"There is much less information about light phenomena. They are seen only in 17%. This is explained by the fact that the weather was clear in the south of the captured space — in the north it was cloudy, firstly, and secondly, for a good half of all observations, the meteor obviously flew close to the Sun, so it was accessible to the observer only in the eastern part of the region. All 17% of the observers of light phenomena were in the eastern part of the region."

In 1997 a new account from Kirensk by Ivan Suvorov (who was in open air) was presented by V.A. Bronshten [Bronshten, 1997]. Despite that the account is from "second-hands", Dr. Vitaly Bronshten was known as a careful fact-checker for his own works, so if he decided to publish some fact then this means that he considers it as reliable. The event took place in the town of Kirensk. Here is how Bronshten describes the event [Bronshten, 1997] (translated):

"Ivan liked to get up early and do jogs in one verst. June 30, 1908 morning was not an exception. This morning was cloudless, the sun brightly shone, no any wind. Suddenly Ivan's attention was drawn by the amplifying noise proceeding as it seemed to him, from southeast side of the sky. Neither from the East, nor from the North, nor from the West nothing similar was felt. The sound came nearer. "All this began. - Ivan Suvorov wrote, - on my watches verified the day before by post-office of Kirensk, at 6 hours 58 minutes local time. Gradually coming source of noise began to be heard from South-South-West side and passed into the West-North-western direction that coincided with the shot-up fiery column up at 7 hours 15 minutes in the morning".

Ivan Suvorov made this record on fields of the illustrated Bible which used in a family. In 1929 - 1930 when Komsomol members - atheists started walking homes and to withdraw religious literature, Agrippina

Vasilyevna herself threw the precious Bible into fire. So Ivan Suvorov's records died.

And nevertheless they were not gone - they remained in memory of his son, Konstantin Suvorov many times reading the story of the father and then restored it.

<…>

What surprises us in these accounts? First of all, time of the beginning of audibility of an abnormal sound - 6 hours 58 minutes while the fiery column shot up, in full consent with other definitions, at 7 hours 15 minutes. The Tunguska bolide could not fly, making a sound, for 17 minutes. During this time at a speed of 30 km/s it would fly by 30000 km, that is at 6 hours 58 minutes it was far outside the atmosphere and could not make any sounds. It means, this moment belongs not to the beginning of emergence of the sound, and to some other event, for example to Ivan's exit from the house.

The correct indication of the moment of explosion forces us to reject also all other possible assumptions: for example, that watches of Ivan lagged behind per day for 17 minutes or that local time of Kirensk strongly differed from local times of other points. Moreover, - in the same Kirensk the director of a meteorological station G. K. Kulesh recorded according to indications of a barograph arrival of an air wave (i.e. the same sounds) after 7 hours.

So inexact Ivan defined and the direction from where the sounds came. The Tunguska bolide flew by, by the most exact definitions, to the North from Kirensk. The closest point of a trajectory was from it to the northeast. Then the bolide moved to the North and, at last, to the northwest.

According to Ye. L. Krinov in his book "Tunguska Meteorite" (Moscow: USSR Academy of Sciences, 1949, p. 54) many eyewitnesses later claimed that they heard the sound before they saw the bolide (which in fact could not be). Apparently, this is some kind of property of inexperienced observers who reported what they saw much later, several years after the event."

The Bronsten's attempts to "improve" the account will be not commented, just adding that a barograph in Kirensk recorded arrival of the air-pressure disturbance at 7 h. 48 min.[Astapowitsch, 1940], and the barograph detected air-pressure disturbances of very low frequency.

Krinov indeed noticed existence of these precursors in accounts and wrote [Krinov, 1949] (translated):

"… , attention is also drawn to the indication of the sounds that preceded the appearance of the bolide. It turns out that this strange feature is noted by many witnesses, independently of each other."

Krinov admitted that the precursors which manifested as "thunders of considerable force" may be a special type of electrophonic phenomena caused by a large scale of the event, or by eyewitness's confusion [Krinov, 1949]. The first explanation is unlikely as can be seen on example of the 2013 Chelyabinsk Meteoritic Event – see [Ol'khovatov, 2021a]. The second Krinov's idea is universal explanation which can explain almost everything. However it should be remembered that the idea of Tunguska as a spacebody infall is based on eyewitness's accounts.

And after the "second-hands" account there is the first-hand account from rather reliable source (and not presented in [Vasil'ev, et al., 1981]). Here is how it is presented by I.S. Astapovich [Astapowitsch, 1940]:

"At Kezhma on the Angara (there is still another, Upper Kezhma, also on the Angara, at Bratsk), A. K. Korin adds to the remarks on the meteorological observations for June 30, 1908 (archives of the former Irkutsk magnetic and meteorological observatory): "At 7 a.m., there appeared in the north two fiery circles of colossal dimensions; $4^m$ after their appearance, the circles vanished; soon after the disappearance of the fiery circles, was heard a powerful noise, resembling the sound of wind, which went from north to south; the noise lasted about $5^m$. Then followed sounds and crashes resembling the discharges of huge cannon, which made the windows rattle. These shots continued for $2^m$ and then was heard a crackling, resembling rifle shots. These latter continued for $2^m$. All this happened when there was a clear sky.""

Unfortunately there is a mistype in the observer's surname – his surname was Kokorin. The observer of the Kezhma meteorological station - Anfinogen Kesarevich Kokorin (1872 – 1938) was interested in the event, collected info, etc. In the late 1920s he even helped to L.A. Kulik (Kulik wrote in 1930 in a letter that Kokorin was considered by local authorities as a valuable employee, but was "lishenets" – i.e. deprived of some civil rights). Unfortunately during the Stalin's terror, Kokorin was accused of counter-revolutionary propaganda and shot. He was rehabilitated in 1958.

As his account is rather important, here is another translation (by A.O.):

"June 30…new calendar at 7 o'clock in the morning, two huge fiery circles appeared in the north; after 4 minutes from the beginning of the appearance, the circles disappeared; soon after the disappearance of the fiery circles, a strong noise was heard, similar to the noise of the wind, which went from north to south; the noise lasted about 5 minutes. Then followed the sounds and crackling, similar to the shots from huge guns, from which the frames trembled. These shots lasted for 2 minutes, and after them there was a crackling sound, similar to a shot from a gun. These last lasted 2 min. Everything that happened was under a clear sky".

It can be seen that the Kokorin's info about luminous phenomena is unique. A possible reason for this will be discussed below.

And finally an official report from ([Vasil'ev, et al., 1981], translated):

"The archives of A.V. Voznesenskii preserved "A copy of the report of the Yenisei County the police Officer to Mr. Yenisei Governor. Dated July 19 , 1908

On the 17th of last June, at 7 o'clock in the morning over the village of Kezhemskoye (on the Angara) from the south towards the north, in clear weather, a huge aerolite flew high in the sky, which discharged, produced a series of sounds like gunshots, and then disappeared. I am informing Your Excellency about this.

County police officer
Solonina.""

## 3. Discussion

In 1925 A.V. Voznesenskii [Voznesenskii, 1925] proposed a trajectory of the Tunguska spacebody basing on accounts collected in 1908. On Fig.1 it is marked as "1". L.A. Kulik and I.S. Astapovich had similar opinion. The average azimuth of the trajectory was found to be 192° (measured from north to east). Then in 1949 Ye. L. Krinov proposed (basing on accounts) his trajectory [Krinov, 1949] which on Fig.1 is marked as "2". Its azimuth is about 137°. In 1955 N.N. Sytinskaya conducted research, but failed to choose between these two trajectories. To calculate preferable azimuth some accounts were claimed as unreliable, while some others as reliable. So every claimed calculated azimuth had its own group of such accounts.

In 1960s two events occurred:  a) an axis of symmetry of the Kulikovskii forest fall was discovered giving azimuth about 115°; b) New accounts were obtained from villages along Nizhnyaya Tunguska river where some luminous body was seen high in the sky (on Fig.1 it is an area where the river flows almost along 108° E longitude). So some new azimuths of the trajectories (basing on accounts) were proposed with azimuths varying from about 126° to 115°.   Later, however, the more accurate azimuth (determined from the fallen trees in the forestfall) was obtained 99°, and azimuth determined from the tree's burn – 95° (see trajectory marked "3" on Fig.1). There are other problems with accounts collected along Nizhnyaya Tunguska:

a) many eyewitnesses claimed that the event happened at about the lunch time.

b) There is a contradiction (within the framework of the meteorite interpretation) which follows from the accounts of eyewitnesses that the angle of inclination of the trajectory of the "Tunguska meteorite" should be much smaller than it turns out in the models of numerical calculation of its fall (~ 35° - 45°). Indeed according to the currently popular trajectory-3, it is assumed that the "Tunguska meteorite" flew near the village of Preobrazhenka. According to V.A. Bronshten [Bronshten,1999]:

"Thus, if it is assumed that the Tunguska bolide was observed at the zenith of Preobrazhenka (s = 350 km), then the inclination of its trajectory was ~15°."

To eliminate this contradiction Bronshten considers the following [Bronshten, 1999]:

"There is, however, another factor that is able to increase the inclination angle of the trajectory at its final position. This is a nonzero aerodynamic lift experienced by the body. Khokhryakov (1977) was the first to note its role in the problem under discussion. Detailed calculations of the effect of the aerodynamic lift (i.e., of the shape) of the body on its incidence angle were carried out by Korobeinikov et al. (1982, 1984). These studies showed that, for certain values of the parameters of the body, the influence of the aerodynamic lift can be quite substantial."

The nonzero aerodynamic lift is associated with powerful energy deposition, with aerodynamic overloads, etc. The question how the alleged Tunguska spacebody could tolerate this "lift" is open…

c) There is one more contradiction which even the "nonzero aerodynamic lift" can't resolve. In the article by A. A. Yavnel [Yavnel, 1988] it is calculated at what angular height (above the local horizon ) the alleged Tunguska bolide (flying on the trajectory-3 with azimuth 99°) can be seen from Kamenskoe. Yavnel got the following (translated):

"As can be seen from the graph (see Fig. 2), the magnitude of the angle ..., above which the bolide could not be observed at all, is 4.4°. This

value is the upper limit of the angular height of the bolide for the most distant observers."

The author of this paper drew a scheme with the Yavnel's calculations (as seen from Kamenskoe) on Fig.5 for better understanding.

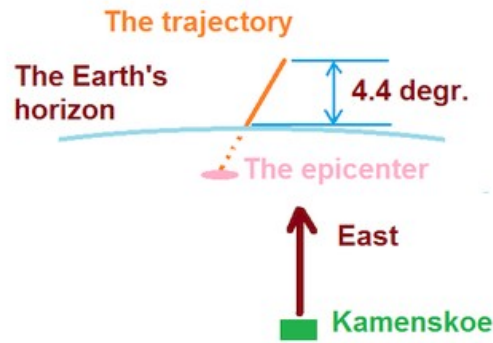

**Fig.5**

In other words, according to Yavnel the alleged Tunguska bolide could be seen very low above the local horizon in Kamenskoe and then swiftly "dived" below the local horizon. However Yavnel did not take into account local (near Kamenskoe) topography. If to look at a map, then it can be seen that there are hills to the east of the village, which completely shade the trajectory-3 view (the map can be seen at https://nakarte.me/#m=13/58.34779/92.49261&l=O/J , also a recent photo of Kamensk –modern name for Kamenskoe, can be seen at: http://temples.ru/show_picture.php?PictureID=43733 ). So residents of Kamenskoe did not see the Tunguska spacebody-bolide moving on trajectory-3, but saw something else. There are also other arguments in favor of this statement, which will be discussed later in this paper.

Academician (of USSR Academy of Medical Sciences) Nikolai Vasil'ev, who was informal leader of the Tunguska research wrote [Vasil'ev, 1992] (a mistype is corrected):

"Analysis of the catalog of statements by eyewitnesses to the disaster [11], the total number of which runs to a few hundred, reveals a fact that has not been clarified to date, namely that thunderlike sounds were heard not only during and after the flight of the bolide, but even before it. <...> It would hardly be realistic to explain them away as subjective errors, since claims of this kind are made over and over and independently of each other.
<...>

    The second factor, a fairly odd factor, is related to the direction of motion of the body. Analysis of statements by witnesses who gathered along the hot tracks of the event [11] and in the 1920s and 1930s [25, 28] led the first investigators of the problem (L. A. Kulik, I. S. Astapovich, and E. L. Krinov) to the unanimous conclusion that the bolide traveled in the direction from south to north. However, analysis of the vector structure of the timber fall due to the shock wave of the Tunguska meteorite gives an azimuth of 114° [29, 30], and the field of burn damage even gives an azimuth of 95° [6-8], i.e., it indicates that the meteorite traveled from nearly east to west. It should be added that this direction also is confirmed by an analysis of the statements of eyewitnesses who lived at the time of the event in the upper reaches of the Lower Tunguska River (in the region of Preobrazhenka, Erbogachen, and Nepa). "

In 1994 Vasil'ev presented more details [Vasilyev, 1994] (TSB is the Tunguska spacebody):

"The first investigators of the Tunguska meteorite (L.A.Kulik, E.L.Krinov, and I.S.Astapovich [1; 2; 3 ]) who analyzed comparatively fresh evidences of the flight of the TSB on the Angara river did not doubt that it had moved generally from the south to the north, though there were three versions of its trajectory (the southern one, proposed by L.A.Kulik, the south-eastern by E.L.Krinov and the south-western by I.S.Astapovich). By the early 60-s it was Krinov's trajectory, namely 135° east of the true meridian, that was considered the most realistic.

Later however, as more information was accumulated on the vector structure of the fallen forest field [9; 17; 59], a "corridor" of axially symmetric deviations of the vectors of the forest falling from the dominating radial pattern was revealed, and this deviation was interpreted as the track of the ballistic wave. The direction of, the "corridor" which was initially estimated as 111° E from N (114° east of the true meridian) [17] was later found to be 95° E from N (99° east of the true meridian) [10], which roughly coincides with the axis of symmetry of the radiant burn area [19]. In this period of time, V. G.Konenkin [60] and later other investigators [61-63] questioned old residents of the area who had lived in the upper reaches of the Nizhnyaya (Lower) Tunguska in 1908 (where there was no questioning in the 20s and 30s). This resulted in the conclusion that TSB had been observed in the said area as well, the analysis of the data suggesting that the body moved from the ESE to the WNW, i.e. by the path coinciding with the projection of that of the TSB, as found on the basis of analysis of the vector picture of the fallen forest area. This coincidence caused revision of the notion of the TSB path, and since the year 1965 the ESE-WNW (in fact, even E-W) version has been accepted in literature. For some years it was assumed to be finally true.

A grave disadvantage of the calculations of TSB path before the mid-80s was that there were analyzed only some separate groups of eye-witnesses' accounts obtained by different researchers, in different periods of time, and not the whole body of evidence. Publication of the catalogue of eye-witnesses information [4] enabled analysis of the whole event. This was done in Ref. [56] and corroborated the considerations expressed earlier in Ref. [58] and also by I.S.Astapovich [64]. Two fundamental facts were established in particular:

1. The total combination of evidence given by "eye-witnesses of the Tunguska fall" contains in fact information on at least two (most likely more) large day-time bolides. It is important that the "images" of the "Angara" and the "Nizhnyaya Tunguska" bolides are quite different and everything seems to indicate that they belong to different objects.

2. The trajectory calculated on the basis of evidences of witnesses of the "Angara" phenomenon and corresponding most likely to its version proposed by E.L.Krinov [1] deviates considerably from that determined by analyzing of the vector structure of the forest fall area and the radiant burn area [9; 19]. Indeed, evidences of the Angara eye-witnesses, including the report of a district police officer, strongly suggest that the bolide flew "high in the sky", which is hardly consistent with the path 99° E of the true meridian. On the contrary, the data obtained on the Nizhnyaya Tunguska river, though agreeing with the configuration of the destruction area, are in contrast with the Angara observations.

An extra complication is that Nizhnyaya Tunguska data suggest virtually unambiguously that bolide's flight took place in the afternoon, unlike those of the Angara which refer to the early morning.

Attempts to resolve the conflict between the data face with considerable problems. If the Angara and Nizhnyaya Tunguska observations are due to different bolides, which is most probably so, then with which of them the destruction area originally explored by L.A.Kulik is associated? Judging by the destruction area configuration, the most probable candidate is the eastern (Nizhnyaya Tunguska) bolide. However none of TSB investigators doubts that the explosion at the distance of 70 km from Vanavara occurred in the early hours of the day, not past midday [56]. Moreover, there is no direct proof that the Nizhnyaya Tunguska bolide was observed in the year 1908, inasmuch as this event was not recorded in any official documents, unlike the Angara bolide.

Besides, even assuming the area of the leveled forest, discovered by L.A.Kulik, to be due to the Nizhnyaya Tunguska bolide, it remains unclear where the Angara bolide fell, then. Throughout the Tunguska "meteorite" study there was no doubt the latter had in fact exploded in the Vanavara region...

But if the forest leveling was caused by the Angara bolide, how does it fit the direction of the "corridor" impressed in the area of the fallen forest by the TSB ballistic wave?"

Vasil'ev underlines the following [Vasilyev, 1994]:

"In the search of way out of this maze, more than one approach has been tried. Some researchers, preferring direct physical evidence, practically ignored eye-witnesses' testimonies as an unreliable subjective material. This approach could be agreed with to some extent, if it were a matter of a few inconsistent testimonies, not many hundreds of independent reports. Besides - what is very important - the testimonies of the year 1908 include official documents of the time, whose authors were responsible to the authorities for their trustworthiness. For this reason, the eye-witnesses' reports should be regarded as a material equal to other data sets or at any rate not to be ignored, even if they do not conform to some speculative arguments."

It is interesting to add that since 1948 several accounts appeared, hinting at the trajectory almost "from west-to east" (not shown on Fig.1), i.e. as being reversed to the trajectory-3 [Yastrebov, 2022; Ol'khovatov, 2020].

In the mid-1960s V.G. Fesenkov (academician of Academy of Science of USSR, and the chairman of the Committee on Meteorites) tried to calculate the "Tunguska spacebody" trajectory basing on the eyewitness accounts collected in 1908, and he had to wrote such a phrase [Fesenkov, 1966] (translated):

"These are rather uncertain conclusions that may be inferred from a review of the most trustworthy eyewitness accounts of 1908."

It is remarkable that interpretation of the Tunguska event as a spacebody infall (which initially mainly appeared due to some eyewitness accounts in preliminary reports) has large problem to conform even to the 1908-accounts in reality…

The Fesenkov's failure to get a reliable trajectory hints that possibly the Tunguska event was associated with more than one luminous body/phenomenon. Indeed, according to the spacebody infall interpretation accounts must match a model of a single spacebody moving on a ballistic trajectory. But it fails… So let's try to understand in detail – why it fails. At first let's look at the 1908-accounts attentively. In the author's opinion it is possible to select several types of luminous objects seen (a reader of this paper is welcome to find even more):
   - 1. two huge fiery circles;
   - 2. strip-like: a pillar of fire; a fiery strip; a body (in the form of a "pipe"), shining with a white bluish light, moving for 10 minutes from top to bottom; etc.;
   - 3. fiery balls;

Let's consider them in more detail.

**1. Two huge fiery circles**. They are described in the Kezhma meteorological station weather log (see early in this paper). They occurred at 7 a.m. when it is time to conduct meteorological observations at the station. The observations included observation of the sky condition (cloudiness, etc.), so must be conducted in open space. Probably that's why the circles were noticed. It is interesting that in the Kezhma meteorological station weather log daily data cloudiness at 7 a.m. is marked as 4, and at 1 p.m. as 10. In other words since appearance of the 'huge fiery circles' a sharp increase of cloudiness took place. The author of this paper inclines to think that there were some relation with the Tunguska event, but his argumentation is beyond the scope of this paper.

It is possible that the circles were one of the first 'luminous' stages of the phenomenon manifestations (at least if looking from remote areas). Some hints on existence of such luminous phenomena are in account obtained in 1924. Engineer V. P. Gundobin informed L.A. Kulik on January 19, 1924 among other things, about the account by Ivan Vasilyevich Kokorin ([Vasil'ev, et al., 1981], translated, a brief 'geographical' remark by Vasil'ev et al. is omitted):

"He, at that time, was traveling in a boat on the Angara. This was at the Mursk rapids at 5 o'clock in the morning of June 17, 1908. He was sailing on the Angara river (with boats). He was the steersman of the boat. His impressions were these: In the north flashed up a bluish light, and a fiery body passed in the sky from the south (considerably larger than the sun), which left behind it a wide light strip; then there occurred such a cannonade that all the workers that were in the boat rushed to hide themselves in the cabin, having forgotten about the danger which threatened from the rapids. The first thunders were weaker, but they increased in strength. The sound phenomena lasted, he estimated, from three to five minutes. The strength of the sounds was so great that the boatmen were entirely demoralized and it took great efforts to return them to their places on the boat."

Interestingly that A.K. Kokorin did not reported about any fiery balls, etc., which according some other reports occurred some minutes after 7 a.m.. The reason for this is not clear. One of possible explanations could be that these luminous phenomena being reported (by other persons) from Kezhma (and in some other places) were at low-altitude (see also below), so just a few people saw them. By the way, it is remarkable that the fiery body (seen by Ivan Vasilyevich Kokorin) was not very bright while having large angular dimensions.

It is interesting that many Evenks who were in the vicinity of the epicenter reported (in the 1960s) redness. So Elkina Anna Yakovlevna, who was at the time of the event about 130 km northeast of the epicenter, said ([Vasil'ev et al., 1981], translated): "The whole sky was red, and not only the sky, everything around was red - the earth and the sky." Then this redness suddenly disappeared. Could this redness be associated with the huge fiery circles? Who knows…

**2. Strip-like.** A pillar of fire was reported from Kirensk (see early in the paper). If to assume that it was positioned in the epicenter, then it means that its height was about 80 km at least. Moreover according to Olga G. Gladysheva [Gladysheva, 2022]:

“Research carried out at the epicenter showed that the Tunguska catastrophe took place at the site of an ancient volcano. However, the greatest surprise was caused by the fact that the main channel for the release of radiation energy turned out to be spatially connected with the central channel of this volcano (Gladysheva and Popov, 2016). <…> If at an altitude of ~2 km above the paleovolcano the outlines of the luminous region are rather vague, then at an altitude of ~6 km the region of maximum luminescence is located exactly above the central vent of the volcano. This is unlikely to be a coincidence.”

Here is what geologists add [Sapronov, et al., 2001](translated):

"The area of the Tunguska meteorite fall is a node of deep faults of the north-western, north-eastern and submeridional directions.
<…>
Thus, geology explains many anomalous phenomena in the area of the Tunguska meteorite fall: the disturbed nature of the magnetic field, various geochemical and gas anomalies, halos of mechanical scattering of many minerals, irregularities of the radiation background, etc."

Remarkably what is written in the book [Yeromenko, 1990] about one of the prominent faults in the region (translated):

"The Berezovsko-Vanavarskii fault (BC, fig. 15, 16) within the Siberian platform was traced from the mouth of the Bakhta river to the settlements of Poligus and Mutorai and further along the route of the Tunguska meteorite in the upper reaches of the Vanavara river."

Anyway it is interesting to compare description of the pillar of fire with what was seen in the village Nizhne-Karelinskoe (see above), which was about 28 km to NW from Kirensk:

"…the peasants saw in the northwest, quite high above the horizon, some extremely strong (it was impossible to look at) a body shining with a white bluish light, moving for 10 minutes from top to bottom. The body was represented in the form of a "pipe", i.e. cylindrical. The sky was cloudless, only low above the horizon, in the same direction in which the luminous body was observed, a small dark cloud was noticeable. It was hot and dry. Approaching the ground (a forest), the shiny body seemed to spread out, a huge puff of black smoke formed in its place and an extremely strong knock (not thunder) was heard, as if from large falling stones or cannon fire. All the buildings were shaking. At the same time, a flame of indeterminate shape began to burst out of the cloud."

It is very interesting that: "…low above the horizon, in the same direction in which the luminous body was observed, a small dark cloud was noticeable." It looks like already there were some events much lower below this 'slow-descending luminous body'. For example, in Nizhne-Karelinskoe the 'slow-descending luminous body' was blindingly bright, while it was not reported such in Kirensk (Kirensk is farther from the epicenter for just about two dozen kilometers, so the difference in distance is unlikely to explain). Why is the difference in description? If to admit that the 'slow-descending luminous body' was not far from Nizhne-Karelinskoe, then what about the pillar/spear seen in Kirensk? Interestingly that from Kezhma reported about a 'flame'- could it be the spear-like (i.e. apparently 'tall-and-thin') Kirensk pillar? It seems unlikely. Indeed, S.B. Semenov who was in Vanavara in 1908 (interviewed by Ye.L. Krinov in 1930) said ([Vasil'ev et al., 1981], translated) that exactly in the north the sky split and a fire appeared widely and high above the forest (as shown by Semenov at an altitude/angle of about 50°). The sky expanded into a large space, the entire northern part of the sky was covered with fire. Vanavara is 65 km from the epicenter, so the flame had significant dimensions horizontally. So questions, questions, questions…

At 2 p.m. between Kirensk and N-Karelinskoe (closer to Kirensk) there was an ordinary thunderstorm with heavy rain and hail. Also as it was shown in [Ol'khovatov, 2021b] some peculiar phenomena took place several dozens kilometers to the west from Kirensk in the first half of June 30. So the question about the 'slow-descending luminous body' is still open…

Now let's search for possible 'pillar of fire' (reported from Kirensk) in the account by Vakulin - a head of the Nizhne-Ilimskoe post-office (see early):

"…according to the stories of a large circle of local residents, they were initially seen in the north-west direction a fireball descending indirectly along the horizon from east to west, which, when approaching the earth, turned into a pillar of fire and instantly disappeared; after disappearing in this direction, a puff of smoke was visible rising up from the earth. A few minutes later there was a strong noise…"

"…turned into a pillar of fire…" – very intriguing… Another account from Nizhne-Ilimskoe (see early) is even more intriguing:

"At the end of the questionnaire M.R. Romanov wrote: "At the beginning of the 9th hour of the morning local time, a fiery ball appeared which flew in the direction from the southeast to the northwest. This ball, approaching the earth, took the form of a flattened ball from above and below (as it was visible to the eye); approaching even closer to the earth, this ball had the appearance of two pillars of fire. When this huge mass fell to the ground, two strong, thunder-like impacts occurred,…"

A prominent geologist S.V. Obruchev wrote in 1925 about his visit to the Tunguska event region in 1924 that ([Vasil'ev, et al., 1981], translated):

"In the Teterya trading post, fire pillars were seen in the north.

It points that there were more 'fire pillars'. It hints that at least some pillars were of relatively low height as they were not reported from Kezhma.

Regarding the strip seen from Znamenskoe approximately to the southwest, it can be said that it does not conform to the idea of the alleged Tunguska spacebody flying on the trajectory-3. Trajectories-1, and -2 can't be ruled out completely, but are unlikely due to the described rather tangible concussions (see also [Ol'khovatov, 2022]).

**3. Fiery balls.** In the 1908-reports they are often mentioned (and partly already are considered above). But at first let's return to the account from Kamenskoe (see early in this paper):

"…had seen a body (as if detached from the sun) more than a yard long, oblong in shape and tapering towards one end; his head was as light as the sun, and the rest was of more hazy color. This body, having flown through space, fell in the northeast."

If the witnesses could not see the alleged Tunguska spacebody on the trajectory-3, then maybe the alleged spacebody flew on the trajectory-1, for example? The answer is: unlikely. And that's why. Azimuth from Kamenskoe to the epicenter is about 58°, and a part of the trajectory-1 would be below hills (looking from Kamenskoe). So the movement of the alleged Tunguska bolide could take place in a spatially (angular) limited area. This small area hardly fit with the words of eyewitnesses: "having flown through space." By the way, please, pay attention, the eyewitnesses from Kamenskoe, being at a distance ~600 kilometers from the trajectory-1 of the alleged 'Tunguska bolide', and looking at him against the Sun, were able to see many details of its structure! And in any case the alleged Tunguska bolide can't be seen as "fell in the northeast" on the trajectory-1 due to hills.

But there are even stronger arguments.

a) According to the account about the sound: "…three underground thunderclaps were heard in the direction from northwest… At the same time, some observed a concussion." The account points to an underground source, and moreover from another direction.

b) A sound generated by the alleged spacebody infall could reach Kamenskoe in about a half an hour - seismic waves from the epicenter (in the frame of the spacebody-infall interpretation) were too weak to produce the sound. But according to the account the sound's delay was just several minutes. Indeed, in the settlement Bel'skoe (Bel'skoe is on almost the same azimuth from the epicenter as Kamenskoe) which is ~60 kilometers from Kamenskoe no sound was heard, just oscillations of the ground [Ol'khovatov, 2022]. And nobody from this relatively large settlement (at the beginning of the year 1911 there were 756 residents) saw any bolide, despite absence of shading hills. Moreover in the town of Yeniseysk (population about ~10 thousand) which is just about 20 km at azimuth ~306° from Kamenskoe thunderclaps were heard, and in some houses windows rattled and lamps waved (report by Sal'nikov M.A. to L.A. Kulik in 1928 [Vasil'ev, et al., 1981], translated). And no any info about any luminous object, despite its population with relatively large police staff (authorities were interested to find out about the event as well as general public), and absence of shading hills.

So, here is what can be said regarding the luminous body seen in Kamenskoe:
1) The body did not fly on any of the trajectories (-1, -2, -3).
2) It can't be the alleged Tunguska spacebody-bolide. Moreover it was hardly caused by a meteoroid. The reasons are following: a) it flew not far from Kamenskoe and apparently on relatively low altitude. As it was not seen from Yeniseysk, then probably its altitude was no large than several kilometers (sooner even no higher than ~1 km), which is hardly possible for a meteoroid. b) The luminous body was associated with powerful acoustic (from underground – according to witnesses) and even seismic phenomena. The latter resembles account from Shamanskie water-measuring posts, where:

"…As local peasants who were at field work at that time told me, they saw some kind of a fiery ball flying in the north side, from which such strong thunders like explosions seemed to occur."

Physical mechanism of such phenomena is not clear still.

It can be added that the body had no any trail as expected from a superbolide. The trail can be small if a meteoroid flies above about 40 – 60 km altitude, but at lower altitudes a trail due to intensive ablation is presented. A good example of a spectacular long-living 'smoke' trail is during the 2013 Chelyabinsk Meteoritic Event. And the alleged Tunguska spacebody is expected to be even much larger than the 2013 Chelyabinsk meteoroid.

The situation is a typical for the Tunguska event: a fiery ball can be seen only in one settlement, or even just in a place close to a settlement. A possible reason for this is a low altitude of the fiery ball.

And now let's look at the most remote place from which luminous phenomena were reported. It is the village Malyshevka which is 798.5 km from the epicenter at azimuth 173°. Remarkably that the 'stump' and the sound appeared simultaneously. In the frame of the Tunguska spacebody infall' interpretation it couldn't be explained as 'electrophonic sounds' on an unprecedented long distance, because the sound was accompanied with some concussions of the ground. If even to propose some small difference between the sound and the 'stump', then it does not help as sound from the alleged Tunguska spacebody demands several tens of minutes to reach Malyshevka from trajectories -2 and -3, and the trajectory-1 can be ruled out due to another direction.

So the 'stump' could not be the alleged Tunguska spacebody bolide. This case resembles the case reported from the Shamanskie water-measuring posts (the fiery ball as a source of the thunders). It is interesting to note that the 'stump' was not point-source of light but had relatively large angular dimensions. Also its brightness was rather moderate. In some way it resembles an object seen during the 1995 Kobe earthquake [Kamogawa, et al., 2005]:

"In Osaka Bay, many fishermen had been working before the main shock. They also saw an orange luminous object moving from the edge of Awaji Island toward Mt. Rokko. Since the luminous object was of an observable size from their locations (approximately 40 km distance) and close to the sea surface in their reports, the diameter and the height would be estimated to be around 100 m. They additionally claimed that the luminous object finally hit Mt. Rokko, causing lightning to strike from the sky toward the ground. Almost immediately after that, they also felt the sea surface moving up and down."

It is interesting to note that about June 30, 1908 there was an upsurge of tectonic activity in the Tunguska event region [Ol'khovatov, 2003] which is generally almost aseismic. For example, on July 5 near Bratskoe there was unusually strong (for the region) earthquake M~3.5 [Ovsyuchenko, et al. 2007].

It can be stated that also (from the 1908-accounts), at least two more luminous objects can be claimed as having nothing to do with the alleged Tunguska spacebody. The first one is which fell to SE of the Voronino (Voronina) village (from the "Sibir" newspaper):

"At that time, in Kirensk, some observed as a fiery red ball in the northwest, moving, according to some, horizontally, and according to others, very obliquely. Near Chechuyskoe, a peasant driving through a field observed the same thing in the northwest.<…>
Near Kirensk in the village of Voronino, peasants saw a fiery ball, fallen to the south-east of them (i.e. in the direction opposite to the one where N-Karelinskoe is located)."

Voronino is very close (2-3 km) to Kirensk (NW of the town). This hints that the 'Voronino' fiery ball was on very low altitude – i.e. the fiery ball could hardly be caused by a meteoroid. Indeed the Kirensk population was several thousand residents

- they hardly could miss the 'Voronino' fiery ball flying at high altitude. Moreover it did not leave any trail. So the 'Voronino' fiery ball could hardly be caused by a meteoroid, and anyway has nothing to do with the alleged Tunguska spacebody.

The second one is from near Ilimsk, where:

"During the thunder, one philistine of Ilimsk was 4 versts from Ilimsk up the Ilim River and saw a "flying star with a fiery tail" that fell into the water, and its tail disappeared into the air."

Of course 'the flying star' had nothing to do with the alleged Tunguska spacebody. Could it be a meteoroid? Ilimsk ceased to exist in 1974 due to flooding by the waters of the reservoir. If you look at a modern map, you can see that in the specified place the river (more precisely, the water reservoir) loops and the width of even the reservoir does not exceed several hundred meters. The shores are hills. Therefore, the distance from the eyewitness to the crash site hardly exceeded several hundred meters. If it was a meteoroid then: a) its speed was at least about 3 km/s to be luminous and to produce its trail. But such speed would result in tremendous effects on the eyewitness which were not reported (by the way, this aspect applies to a number of other cases considered). b) Due to its high speed and proximity to the eyewitness, the meteoroid would most likely have been visible as a streak of light leaving tail. c) It is very unlikely for a relatively small meteoroid to reach the level of the Ilim River with such high speed.

So in this case, it is extremely unlikely that it was a meteoroid.

By the way, from all accounts of 1908 it is the only one which said about a more or less long trail. Early in the paper there was already presented the 1924 account 'with a trail' by Ivan Vasilyevich Kokorin. In the accounts collected in 1920s – 1930s there are several accounts with 'a trail'. Here is the one of them ([Vasil'ev, et al., 1981], translated):

"E.E. Sarychev, interviewed by D.F. Landsberg in Kansk 11/X-1921
"I was a master tanner. In summer (closer to spring) about 8 o'clock (before lunch) I was washing wool with the workers on the bank of the Kan River, when suddenly at first I heard a noise like from the wings of a frightened bird, in the direction from the south to the east, to the village of Ancir, and a wave like a swell {ripples – A.O.} went upstream along the river. After that, one sharp thunder followed, followed by dull, as if underground rumblings. The thunder was so strong that one of the workers, Egor Stepanovich Vlasov (he is now dead), fell into the water. With the appearance of noise, a kind of radiance appeared in the air, circular in shape, about half the size of the moon with a bluish tinge, flying rapidly in the direction from Filimonovo to Irkutsk. Behind the radiance, a trail remained in the form of a bluish strip, stretching almost along the entire path and then gradually disappearing from the end. The radiance, without breaking, disappeared behind the mountain. I could not notice the duration of the phenomenon, but it was very short. The weather was perfectly clear and it was quiet""

As the eyewitness reported about the direction of the 'radiance' flight (as he thought), here it is on the 1910-map (Fig.6).

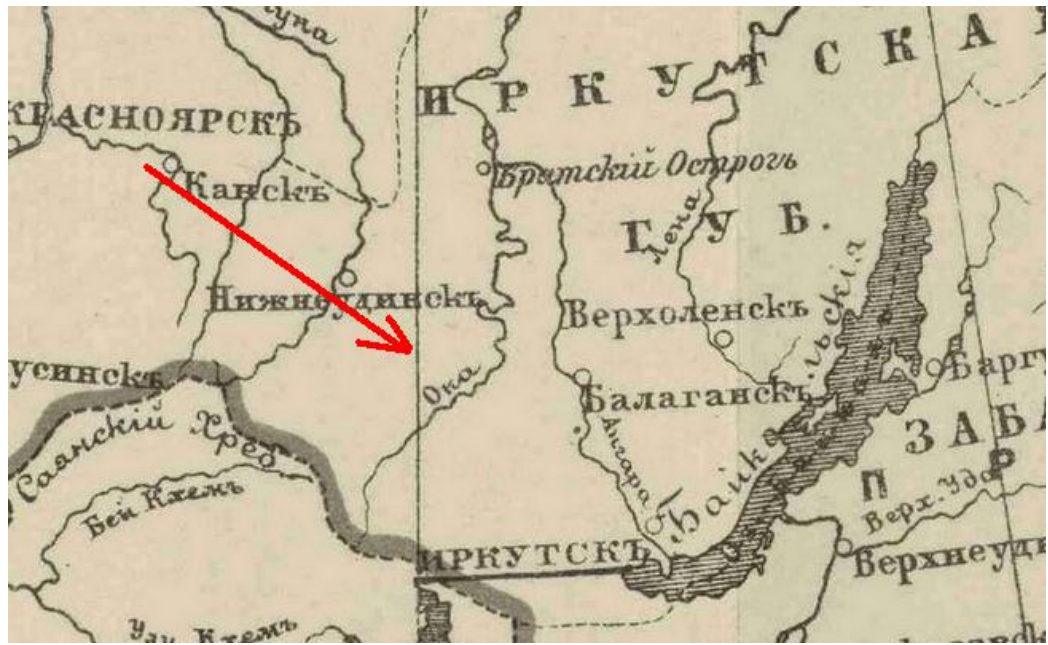

**Fig.6**

It follows from the text, that Sarychev was to the west, rather, even to the southwest, from Ansir. This means that he was about 12-13 km west of Kansk and about 7-8 km north of Filimonovo. There were no any reports of any luminous phenomena from Filimonovo, as well as there were no any reports of any luminous phenomena from Kansk (only sound and seismic phenomena). The Kansk's population was ~ 10 thousand, so unlikely that such 'luminosity' could be missed, unless it was on low altitude. However the story of the Tunguska meteorite started with the newspsper report about a meteorite which fell near Filimonovo almost on the railroad-track. But later no meteorites were discovered, only terrestrial rocks. There are various interpretations of the 'Filimonovo' story, including the 'newspaper fake', etc., but all of them are speculative still. For those who are interested in the history of the Tunguska research here is a fragment of a note from the newspaper "Sibirskaya Zhizn" (published in Tomsk) of July 15, 1908 (translated):

"**About the fall of a meteorite near Kansk**

In one of the issues of "Sib. Zhizn" it was reported that, according to rumors, near the junction Filimonovo Sib. zh. d. {the Siberian railroad - A.O.} a meteorite fell. This rumor was subsequently confirmed and by other Tomsk newspapers. Now we are informed that after appearing in "Sib. Zhizn" of this article, the dean of the mining department of the Tomsk Technological Institute, prof. V. A. Obruchev, sent a telegram to the head of this junction for verification. The latter replied that the meteorite did not fall near Filimonovo, but, according to rumors, near the village of Lovat (Filimonovo is located a few versts west of Kansk, and the village Lovat to the east). Then the laboratory assistant of the Institute, who was in Krasnoyarsk, P. P. Gudkov was authorized to find the meteorite, and, if possible, bring it to Tomsk.

At the request of P. P. Gudkov , the Yeniseisk's governor ordered the Kansk police officer, whose residence is located six versts from the village Lovat to check the rumors about the fall of a meteorite. In response to this, the police officer informed that his personal investigation turned out that there was neither a meteorite fall near Lovat nor in the vicinity of Kansk at all. In view of the fact that on June 17 in Irkutsk province and in some places the Yeniseisk's {province - A.O.} (in the Kansk and the Yeniseisk districts, along the Angara {river - A.O.}, as well as in the mines of the southern taiga), a fairly significant oscillation of the soil was observed, noted by seismographs in Irkutsk, it is possible to attribute the emergence of rumors about a meteorite as a wish to explain this earthquake."

Let's return to the Sarychev account. From the account it is possible to state that the 'radiance' had nothing to do with the alleged Tunguska spacebody. Moreover if it was a meteoroid then: a) its speed was at least about 3 km/s to be luminous and to produce its trail. But such speed would result in tremendous effects on the eyewitness which were not reported (the sound phenomena were simultaneous of the appearance of the 'radiance'). b) Due to its high speed and proximity to the eyewitness, the meteoroid would most likely have been visible as a streak of light leaving a trail. c) It is very unlikely for a relatively small meteoroid to reach the level of the Kan River with such high speed. So in this case, it is extremely unlikely that it was a meteoroid. By the way, of a special interest is the wave (like swell/ripples) which went upstream along the river.

Regarding the considered trails it is possible to say that they sooner resemble something like a fog, than the spectacular long-living smoke trail after the 2013 Chelyabinsk Meteoritic Event.

Another interesting example of such the 'Tunguska' trail is from account by Tikhon Naumovich Naumenko (1886 - 1938). The account was written in Moscow on Jan.21, 1936. Here is translation of its main part from [Vasil'ev, et al., 1981] (at the time under discussion in 1908, Naumenko was a political exile in Kezhma, and took a job helping carpenters to build a barn):

"Definitely I do not remember, on June 17th or 18th, 1908 about 8 o'clock in the morning we with comrade Grabovskii planed boards with a two-handed plane. The day was extremely sunny and so clear that we did not notice any cloudlet on the horizon; wind did not move, - the total silence.

... I sat a back to the Angara River, - to the South, and Grabovskii - facing me.... And here about 8 o'clock in the morning (the sun already rose quite highly) suddenly being-far-far-away, hardly audible sound of a thunder slightly was heard; it forced us to look around involuntarily extensively: at the same time - the sound was heard as from the Angara River so I at once had to turn back abruptly in that side where I sat a back, but as to the horizon in the sky around us any cloudlet was not visible anywhere..., we, believing that a thunderstorm still somewhere away from us, began to plane boards again. But the sound of a thunder began to amplify so quickly that we did not manage to plane more than three-four times, and we had to throw the plane and not to sit any more, and to rise from boards as the thunder sound seemed to us already something unusual as clouds on the horizon were not visible; at the same time, at the moment when I got up from boards, among quickly amplifying sound of a thunder the first, rather weak blow {impact – A.O.} sounded; it forced me to turn quickly a half-turn to the right, i.e. to the southeast from where beams of a bright sun fell on me, and I had to raise eyes a little up in the direction of the heard blow, in that the direction from where beams of the sun watched at me. This complicated a bit observations of that phenomenon which seemed nevertheless seen for an eye at the moment after the first blow, namely: when I quickly turned in the direction of blow, beams of the sun were crossed (across) by a wide fiery white strip on the right side of beams, and with left towards the North (or if to take from Angara then - behind  the Kezhemsky field) an irregular shaped, even more fiery white (paler that the sun, but almost identical with beams of the sun) little oblong body (in the form of a cloudlet, diameter is much more than the moon..., without the correct outlines of edges) flew to taiga.

... After the first weak blow, approximately in two-three seconds, and even maybe more (we had no watches, but the interval was large) - the second, quite strong blow sounded. If to compare it to a thunderclap, then it was the strongest what are during a thunderstorm. After this second blow... the lump {'cloudlet' – A.O.} already disappeared, but the tail (to be more correct - a strip), already all came to be on the left side of beams of the sun, having cut them, and became many times wider, than was on the right side; and right there, through shorter period, than was between the first and second blow, the third thunderclap {blow, impact – A.O.} and such strong (and as if with several blows merged together inside it), even with a crackling followed that all earth began to tremble, and such echo spread over taiga, and even not an echo, but some deafening continuous rumble was carried; it seemed that this rumble captured all taiga of immense Siberia.

It should be noted that the carpenters working at construction of the specified barn after the first and second blows crossed themselves in complete perplexity  (they were 6-7 persons, everybody is local peasant; they were almost all old men then); and when the third blow sounded, so carpenters fell from rishtas {scaffolding – A.O.} on chips backwards (it was low, - one meter and a half), and some were so strongly stunned and frightened that to us with comrade Graboskii it was necessary to bring them round and to calm, saying that everything already passed; but they expected still continuation and said that already likely the end to the World comes and there will be a Last Judgement, etc. They also did not want to listen to our calms, - abandoned work; and we ( it is necessary to admit), were also in perplexity from such unusual phenomenon and as all of us found it difficult to explain an essence of such phenomenon, gave up work too and went to the settlement;... in the settlement there were about 30 more people of political exiled, among them were also with the higher education and therefore we considered that we will receive an exhaustive explanation of this phenomenon from them.

When we came to the settlement, we saw on streets the whole crowds of people ( both locals, and our companions exiled ) who were hotly discussing and on various ways interpreting this unusual phenomenon; because all our companions at the time of flight of a meteorite were in rooms, and some even slept, and they were woken by these unusual force thunderclaps from which even windows ringed ( glasses of windows to be more correct), and in some houses (as also our companions and, especially,  local peasants told) even furnaces cracked and kitchen utensils fell from shelves from severe concussion of the soil; at the same time locals, as well as carpenters working with us, with horror on faces unconsciously interpreted this phenomenon which they never observed earlier, not differently as {i.e. 'only as' – A.O.} superstitious thoughts of the end of the World and approaching "Last Judgement" and so forth heresies.
<…>

"

Naumenko added that almost the whole day passed in different interpretations about this phenomenon among all the villagers. Naumenko listened more to the opinions of his comrades, especially those who had at least an incomplete higher education. In their explanations they made assumptions about the fall to Earth of a rare and unusual meteorite. His account was published in "Meteoritika" in 1941. In the publication also his sketch of the 'flying cloudlet' was printed (see Fig.7).

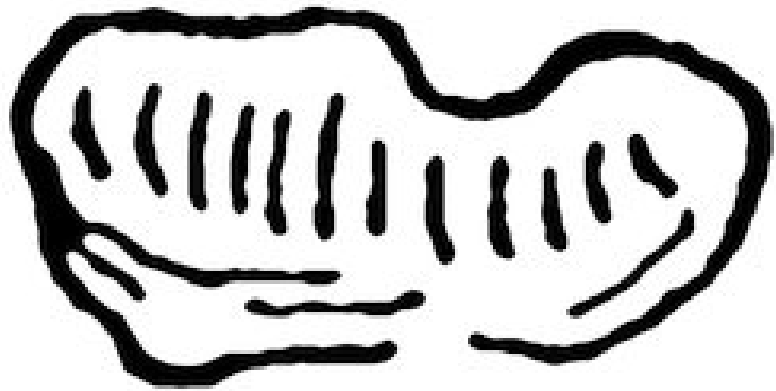

**Fig.7**

So instead of the alleged Tunguska-spacebody superbolide there was a flying cloudlet (with an angular diameter much more than the moon), which soon disappeared… Moreover, its tail was very short: "…but the tail (to be more correct - a strip), already all came to be on the left side of beams of the sun…". Also the tail did not resemble the spectacular smoke trail associated the 2013 Chelyabinsk Meteoritic Event.

Some additional info is presented in the second (electronic) edition of the catalog of the Tunguska event accounts ("Katalog Pokazanii ochevidtsev Tungusskogo padenia") issued in 2018 and deposited at web-page http://tunguska.tsc.ru/ru/science/1/eyewitness/ . Here is from the 2018 catalog (translated):

"In the State Archive of the Tomsk region, in the archive of N.V. Vasiliev, there is photo of the drawing, which is given below, with the inscription on the back: "The drawing on the film was placed between the letters of Naumenko, an observer from Kezhma." We are talking about the photographic film of the KMET {Committee for Meteorites, The Soviet Academy of Sciences – A.O.} Archive.
When typing documents from this film, the drawing was not included

(apparently for a technical reason), but its verbal description was compiled with the preceding inscription: "Naumenko's drawing".

The compilers leave the situation with these drawings without comment."

Here is the verbal description of the "Naumenko's drawing" which is presented on Fig.8 below:

"A circle divided by diameter is drawn, the lower half is drawn with dotted dots, the diameter and the upper half are a solid line. Another dotted line has been added to the lower half of the circle a line, between them the inscription in a semicircle "Silhouette looks like a moon", "lambs" are added to the upper half of the circle along the entire length, an arrow points to the lambs on the top left, under it the inscription "Clouds are also light", on the top right the tail departs from the hemisphere in the form of a broken ribbon (a pennant), gradually tapering to the end and decreasing the swing of the tape to the end (the pennant in the perspective). To the middle of the tape there is an inscription in four lines "The brightness is the same, up to about half of this arrow", under the second half of the tape there is an inscription in two lines "Transition to a dim color, the height is huge". Inside the hemispheres there are inscriptions: in the upper "Brightness is equal to the sun, there were perceptible rays", in the lower "Later this semicircle appeared". Under the picture there is the explanation, "At first, from the ground, the diameter of this ball seemed to be from 180-200 m / m, and then somehow lengthened in the direction of the arrow, moving from a circle to an ellipse.""

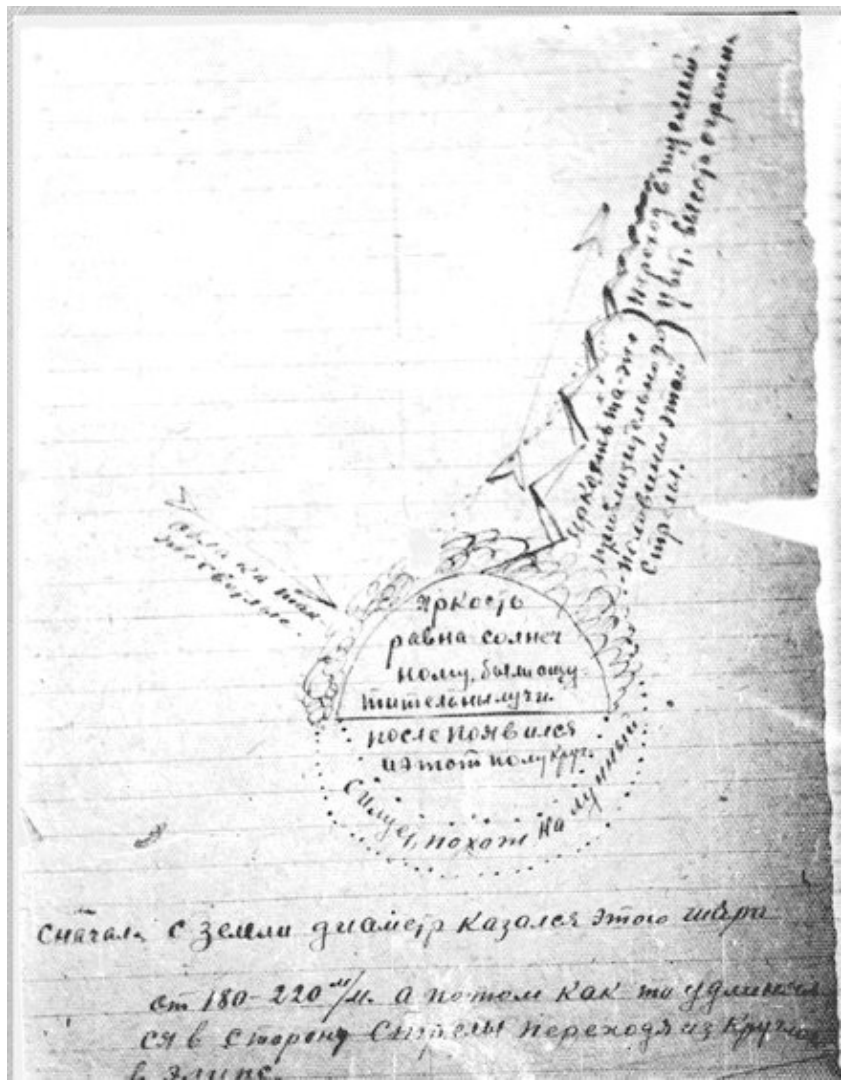

**Fig.8**

Indeed this drawing/sketch shows rather complicated structure of the luminous phenomenon. However there is already something similar: "At first…this ball…, and then somehow lengthened in the direction of the arrow, moving from a circle to an ellipse." It resembles the transformation of a ball into an ellipse (resembling the one which was discussed early in this paper). By the way, in 1984 it was shown that the Naumenko's account is not compatible with the alleged Tunguska spacebody trajectories with azimuth less than 125° [Demin, et al, 1984], i.e. with the trajectory-3 which is widely accepted now in the frame of the Tunguska spacebody infall interpretation.

Naumenko was about 0.5 km from Kezma. Let's look what other eyewitnesses saw from Kezhma. Several accounts were collected in 1930s.

In 1930 K.A. Kokorin was interviewed by Ye.L. Krinov. Here it is ([Vasil'ev, et al., 1981], translated):

"Kokorin does not remember the exact day and year of the fall, but he remembers that three or four days before Petrov's day, at 8-9 o'clock in the morning, no later. The sky was completely clear, there were no clouds. He entered the bathhouse (in the courtyard), managed to take off his top shirt, when suddenly he heard sounds like cannon shots. He immediately ran out into the courtyard, open to the southwest and west. At this time, the sounds were still continuing, and he saw in the southwest, at an altitude of about half the distance between the zenith and the horizon, a flying red ball, and on the sides and behind his rainbow stripes were visible. The ball flew for 3-4 seconds, disappeared in the northeast (the directions were restored from memory when giving a reading of 1/I-1930 by compass). The sounds were heard during the flight of the ball, but they immediately stopped when the ball disappeared behind the forest.<…>"

It is remarkable that the sounds seem to be associated with the ball (by the way the ball does not seem to be bright). Also interesting that K.A. Kokorin saw the ball in the southwest initially, i.e. in the other side of the world than the "cloudlet" that Naumenko saw.

One more account from Kezhma ([Vasil'ev, et al., 1981], translated):

"Bryukhanov A.K. was interviewed in 1929 by the teacher Z. Vostrikova, who gave her notes to L.A. Kulik. Lived in Kezhma.
" ... I haven't had time to get dressed yet after the bath, as I hear a noise. I jumped out into the street as I was and immediately threw a glance at the sky, because I could hear the noise from there. And I see: blue, green, red, hot (orange) stripes are going across the sky, they are as wide as the street.
The stripes went out and the rumble was heard again and the ground shook. Then again and again the stripes appeared and went "under the north". It seemed that they were about 20 versts from Kezhma. Well, then I heard that the end of them was far away, in the Tungus campsite.<…>""

Although Bryukhanov A.K. did not see any ball, his testimony echoes the testimony of Kokorin K.A. in terms of "rainbow stripes". It is noteworthy that their appearance and disappearance were synchronized with the sounds and the earth concussions. It looks like some unknown process rules both the luminous phenomena, and the sounds (and the ground concussions). Apparently it is beyond modern science frontiers still.

And finally from the early Kezhma's accounts ([Vasil'ev, et al., 1981], translated):

> "Bryukhanov D.F., interviewed by L.A. Kulik in 1938, said:
> "At that time I was plowing arable land on Narodnaya (6 km. to the west of Kezhma), when I sat down to breakfast near my plow, suddenly there were blows, like cannon shots. The horse fell to its knees. Flame flew out from the north side over the forest. I thought: the enemy is shooting (at that time they were talking about the war). Then I see the spruce forest bent down: I think about the hurricane, grabbed the plow with both hands so that it wouldn't carry. The wind was so strong that it blew a little soil off the surface of the earth; and then this hurricane drove the water on Angara with a {big -A.O.} wave: I could see everything well, because the arable land was on a hillock.""

Remarkably that the eyewitness saw just the flame. Neither the alleged Tunguska spacebody superbolide, nor even its trail. He did not see the spear-like pillar (which was seen from Kirensk). Also he did not see the 'slow-descending luminous body' which was seen from Nizhne-Karelinskoe.

It follows from the account that the sounds and the flame were simultaneous, or even the sounds preceded the flame.

Regarding possible precursor (sounds preceding the luminous event), here is an interesting account ([Vasil'ev, et al., 1981], translated):

> "A member of the management of the Consumers Association of Kezhma, I. K. Vologzhin, reported the following to L.A. Kulik on the 21st of November, 1921: "In the year 1908, on June 17, I was about 20 versts above the village Kezhma (along the Angara river) in the place Chirida. In the morning we examined our {probably fishing - A.O.} nets. It was a clear morning. There was not a cloud. With me was an old man. We heard - a thunder, then - another one, and gradually the thunder rumblings began to diminish toward the north. I don't remember what time it was, but the sun hadn't risen yet, it was just dawn. In the winter the inhabitants of the village Kezhma, who traded around the Podkamennaya Tunguska...""

It follows from the account that something peculiar took place in the region several hours before the Tunguska explosion.

## 4. Conclusion

A lot of interesting and even intriguing data can be obtained from the 1908-accounts. Unfortunately sometimes reports lack enough detailed data to get a reliable solution about some aspects of the Tunguska event. For example, a nature of the bright 'slow-descending luminous pipe' seen from Nizhne-Karelino is completely unclear.

However it is possible to state that at least some of the reported luminous phenomena have nothing to do with the alleged Tunguska spacebody-bolide. Any interpretation of Tunguska should explain them.

The author hopes that a reader of this paper will make own analysis of the accounts and possibly discover something new regarding this peculiar and unusual Tunguska event. Anyway there are still many aspects of Tunguska need to be researched.

**ACKNOWLEDGEMENTS**
The author wants to thank the many people who helped him to work on this paper, and special gratitude to his mother  - Ol'khovatova Olga Leonidovna (unfortunately she didn't live long enough to see this paper published...), without her moral and other diverse support this paper would hardly have been written.